\documentclass[twocolumn,showpacs,preprintnumbers,amsmath,amssymb]{revtex4}
\usepackage{tipa}
\usepackage{amssymb}
\usepackage{txfonts}
\usepackage{hyperref}
\usepackage{graphicx}
\usepackage{epsfig}
\begin{document}

\title{Reconstructing the Equation of State for Dark Energy In the Double Complex Symmetric Gravitational Theory}
\thanks{Supported by National Science Foundation of China under
Grant No 10573004}
\author{Shao Ying,\footnote{Email: sybb37@student.dlut.edu.cn; Photo code: 0411-84707869(office)}
Gui Yuan-Xing \footnote{Email: thphys@dlut.edu.cn; Photo code:
0411-84706203(office)}, Wang Wei} \affiliation{Department of
Physics, Dalian University of Technology, Dalian 116023}


\begin{abstract}
We propose to study the accelerating expansion of the universe in
the double complex symmetric gravitational theory (DCSGT). The
universe we live in is taken as the real part of the whole
spacetime ${\cal M}^4_C(J)$ which is double complex. By
introducing the spatially flat FRW metric, not only the double
Friedmann Equations but also the two constraint conditions $p_J=0$
and $J^2=1$ are obtained. Furthermore, using parametric $D_L(z)$
ansatz, we reconstruct the $\omega^{'}(z)$ and $V(\phi)$ for dark
energy from real observational data. We find that in the two cases
of $J=i,p_J=0$ and $J=\varepsilon,p_J\neq 0$, the corresponding
equations of state $\omega^{'}(z)$ remain close to -1 at present
($z=0$) and change from below -1 to above -1. The results
illustrate that the whole spacetime, i.e. the double complex
spacetime ${\cal M}^4_C(J)$, may be either ordinary complex
($J=i,p_J=0$) or hyperbolic complex ($J=\varepsilon,p_J\neq 0$).
And the fate of the universe would be Big Rip in the future.
\end{abstract}
\maketitle
\section{INTRODUCTION}
Current astrophysical observations have indicated that the
universe undergoes accelerated expansion during recent redshift
times$^{[1,2]}$. The accelerating expansion has been attributed to
the existence of mysterious dark energy$^{[3]}$ with negative
pressure which can induce repulsive gravity and thus cause
accelerated expansion. The cosmological constant $\Lambda$ with
equation of state $\omega=\frac{p}{\rho}=-1$$^{[4]}$ is the
simplest and most obvious candidate for dark energy. However this
model raises theoretical problems related to the fine tunned
value. These difficulties have led to a variety of alternative
models where the dark energy component varies with time (eg.
quintessence, phantom etc.)$^{[5]}$. Other physically motivated
models predicting late accelerated expansion include modified
gravity$^{[6]}$, chaplygin gas$^{[7]}$, braneworld$^{[8]}$,
quintom$^{[9,10]}$ etc. It is interesting that the equation of
state $\omega$ for the quintom model crosses -1 in the near past.
In addition, some researches have presented to reconstruct
the properties of dark energy from observations$^{[11]}$.\\
\mbox{}\hspace{15pt}On the other hand, differing from these
models, Moffat proposed the nonsymmetric gravitational theory
(NGT)$^{[12]}$ as a possible alternative of dark energy$^{[13]}$.
Meanwhile, a so-called the double complex symmetric gravitational
theory (DCSGT) have been established$^{[14]}$. Moreover, the
corresponding double complex Einstein's field equations have been
obtained. NGT and DCSGT are also the generalized gravitational
theory presented as the unified field theory of gravity and
electromagnetism. Thus, the questions arise whether we can obtain
the corresponding equations of state for dark energy in the DCSGT
or whether the accelerating expansion of the universe can be
studied in the DCSGT. In this paper, we will answer these
questions. For instance, the universe we live in has been taken as
the real part of the whole spacetime ${\cal M}^4_C(J)$ which is
double complex. This paper is organized as follows. We obtain in
Sec two, by introducing spatially flat FRW metric, the double
Friedmann Equations in the DCSGT and the two constraint conditions
$p_J=0$ and $J^2=1$. In Sec three, the equation of state
$\omega^{'}(z)$, potential $V(\phi)$ and scalar field $\phi$ for
dark energy are reconstructed from real observational data. Sec
four is a conclusion.
\section{THE FRIEDMANN EQUATIONS IN THE DCSGT}
In the double complex symmetric gravitational theory (DCSGT),
metric tensor is a double complex symmetric tensor.
Correspondingly, connection and curvature are forced to be double
complex. The real diffeomorphism symmetry of standard Riemannian
geometry is extended to complex diffeomorphism symmetry. In the
double complex manifold of coordinates ${\cal M}^4_C(J)$, the
double complex symmetric metric $g_{\mu\nu}(J)$ is defined by
$$
{g_{\mu\nu}}(J)={s_{\mu\nu}}+J{a_{\mu\nu}},\eqno(1)
$$
where ${s_{\mu\nu}}$ and ${a_{\mu\nu}}$ are the real symmetric
tensors, and the double imaginary unit $J=i,\varepsilon$ ($J=i,
J^2=-1; J=\varepsilon, J^2=1, J\neq 1$)$^{[15,16]}$. The real
contravariant tensor $s^{\mu\nu}$ is associated with $s_{\mu\nu}$
by the relation
$$
s^{\mu\nu}s_{\mu\sigma}=\delta^\nu_\sigma,\eqno(2)
$$
and also
$$
g^{\mu\nu}g_{\mu\sigma}=\delta^\nu_\sigma.\eqno(3)
$$
\mbox{}\hspace{15pt}The double complex symmetric connection $
{\Gamma}_{\mu\nu}^{\lambda}(J)$ and curvature $R_{\mu\nu}(J)$ are
determined by the equations
$$
{g_{\mu\nu;\lambda}(J)}={\partial_\lambda}{g_{\mu\nu}(J)}-{g_{\rho\nu}(J)}{g_{\mu\lambda}^\rho}(J)-{g_{\mu\rho}(J)}{\Gamma_{\nu\lambda}^\rho(J)}=0,\eqno(4)
$$
$$
R_{\mu\nu\sigma}^{\lambda}(J)=-{\partial_\sigma}{{\Gamma}_{\mu\nu}^{\lambda}(J)}+
{\partial_\nu}{{\Gamma}_{\mu\sigma}^{\lambda}(J)}+{{\Gamma}_{\rho\nu}^{\lambda}(J)}{{\Gamma}_{\mu\sigma}^{\rho}(J)}
-{{\Gamma}_{\rho\sigma}^{\lambda}(J)}{{\Gamma}_{\mu\nu}^{\rho}(J)}.\eqno(5)
$$
From curvature tensor, we can obtain the double complex Bianchi
identities
$$
{{\bigg(}{R^{\mu\nu}}(J)-\frac{1}{2}g^{\mu\nu}(J)R(J){\bigg)}_{;\nu}}=0.\eqno(6)
$$
The action is denoted by $S={S_{grav}}+{S_M}$, where $S_{grav}$
and $S_{M}$ are respectively gravity action and matter action
$$
{S_{grav}} =\frac{1}{2}\int{d^4}x{\bigg[}{{\cal
 G}^{\mu\nu}(J)}{R_{\mu\nu}(J)}+{{\big(}{{\cal
 G}^{\mu\nu}(J)}{R_{\mu\nu}(J)}{\big)}^\dag}{\bigg]},\eqno(7)
$$
$$ \frac{1}{\sqrt{-g(J)}}(\frac{\delta
S_M}{\delta g^{\mu\nu}(J)})=8\pi G{T_{\mu\nu}}(J),\eqno(8)
$$
where ${\cal G}^{\mu\nu}(J):=\sqrt{-g(J)}g^{\mu\nu}(J)={\cal
 S}^{\mu\nu}+J{\cal
  A}^{\mu \nu}$, ``$\dag$''denotes complex
 conjugation and $T_{\mu\nu}(J)=\tau_{\mu\nu}+J{\tau_{\mu\nu}^{'}}$ is a
 double complex symmetric source tensor. The variation with respect to
 $g^{\mu\nu}(J)$ and dividing the equation by $\sqrt{-g(J)}$ can lead to
$$
R_{\mu\nu}(J)-\frac{1}{2}{g_{\mu\nu}(J)}R(J)=-8\pi
 G{T_{\mu\nu}(J)}.\eqno(9)
 $$
Eq.(9) is the double complex Einstein's field equation in the
DCSGT$^{14}$. Eq.(9) is written as
$$
R_{\mu\nu}(J)=-8\pi
G\big{(}T_{\mu\nu}(J)-\frac{1}{2}g_{\mu\nu}(J)T(J)\big{)}.
$$
\mbox{}\hspace{15pt}Let us consider the spatially flat FRW metric
$$
{ds^2}=-{dt^2}+{a^2}(t)[{dr^2}+{r^2}({d\theta^2}+{\sin^2}\theta{d\phi^2})].\eqno(10)
$$
The double complex symmetric metric tensor $g_{\mu\nu}(J)$ is
determined by
$$
{g_{00}(J)}=-(1+J),
$$
$$
{g_{11}(J)}=a^2(t)+Ja^2(t),
$$
$$
{g_{22}(J)}=a^2(t)r^2+Ja^2(t)r^2,
$$
$$
{g_{33}(J)}=a^2(t)r^2\sin^2\theta+Ja^2(t)r^2\sin^2\theta.\eqno(11)
$$
The energy-momentum tensor $T_{\mu\nu}(J)$ in the double complex
symmetric spacetime is defined
$$
T^{\mu\nu}(J)=[({\rho_C}+{p_C}){U^\mu}{U^\nu}+{p_C}{g^{\mu\nu}}]+J[({\rho_J}+{p_J}){U^{'\mu}}{U^{'\nu}}+{p_J}{g^{\mu\nu}}],\eqno(12)
$$
where $\rho_C,p_C$ and $\rho_J,p_J$ are energy density and
pressure respectively in real and imaginary spacetime
respectively. We define
$$
s_{\mu\nu}{U^\mu}{U^\nu}=-1,~~~~~a_{\mu\nu}{U^{'\mu}}{U^{'\nu}}=-1,\eqno(13)
$$
$$
T_{\mu\nu}(J)={g_{\mu\alpha}}{g_{\nu\beta}}{T^{\alpha\beta}(J)},\eqno(14)
$$
Substituting Eqs.(11) and (12) into Eq.(9), the double evolutive
equations of the universe are expressed
$$ \frac{\ddot{a}}{a}=-\frac{4\pi
G}{3}[(3{p_C}+{\rho_C})+{J^2}({p_J}-{\rho_J})],\eqno(15)
$$
$$
{p_C}-{\rho_C}+(2-{J^2}){\rho_J}+(4-{J^2}){p_J}=0,\eqno(16)
$$
$$
2\frac{\dot{a}^2}{a^2}+\frac{\ddot{a}}{a}=4\pi
G[({\rho_C}-{p_C})+{J^2}({\rho_J}-{p_J})],\eqno(17)
$$
$$
2\frac{\dot{a}^2}{a^2}+\frac{\ddot{a}}{a}=4\pi
G[({\rho_C}-{p_C})+{J^2}{\rho_J}+({J^2}-2){p_J}],\eqno(18)
$$
where dot means derivative with respect to time.\\
\mbox{}\hspace{15pt}From Eqs.(15)-(18), we can obtain two
constraint conditions
$$
p_J=0~~~~or~~~~J^2=1,\eqno(19)
$$
and Hubble parameter
$$
H^2=\frac{8\pi G}{3}[\rho_C+\frac{1}{2}J^2(\rho_J-p_J)],\eqno(20)
$$
$$
\dot{H}=-4\pi G(\rho_C+p_C).\eqno(21)
$$
Eqs.(20) and (21) are the Friedmann Equations in the DCSGT.\\
\mbox{}\hspace{15pt}In the following section, we will study the
reconstructions of the equations of state for dark energy in the
two constraints $p_J=0$ and $J^2=1$, respectively.
\section{THE RECONSTRUCTIONS OF EQUATIONS OF STATE $\omega_\phi$ AND POTENTIAL $V(\phi)$ FOR DARK ENERGY}
If the density and pressure of the matter of the universe are
respectively
$$
\rho_C=\rho_m+{\rho^{'}},p_C=p_m+p^{'},\eqno(22)
$$
where $\rho_m=\frac{3H^2_0}{8\pi G}{\Omega_{0m}}(1+z)^3$ and
$p_m=0$ are energy density and pressure of nonrelativistic matter,
$\rho^{'}$ and $p^{'}$ are energy density and pressure for scalar
field $\phi$
$$\rho^{'}=\frac{\lambda}{2}\dot{\phi}^2+V(\phi),
p^{'}=\frac{\lambda}{2}\dot{\phi}^2-V(\phi),\eqno(23)$$ where
$\lambda=\pm1$ corresponding to ordinary scalar field and phantom
scalar field. Substituting Eqs.(22) and (23) into Eqs.(20) and
(21), then we can obtain
$$
H^2=\frac{8\pi
G}{3}[\rho_m+\frac{\lambda}{2}\dot{\phi}^2+V(\phi)+\frac{1}{2}J^2(\rho_J-p_J)],\eqno(24)
$$
$$
\dot{H}=-4\pi G(\rho_m+\lambda\dot{\phi}^2).\eqno(25)
$$
\subsection{the case of $p_J=0$}
If the first constraint condition $p_J=0$, Eq.(24) is rewritten as
$$
H^2=\frac{8\pi
G}{3}{\bigg[}\frac{4-J^2}{2(2-J^2)}{\rho_m}+\frac{\lambda}{2}\dot{\phi}^2+\frac{2}{2-J^2}V(\phi){\bigg]}.\eqno(26)
$$
From Eqs.(25) and (26), the double potential $V(z)$ and scalar
field $\phi$ can be obtained
$$
V(z)=\frac{3{H_0}^2}{8\pi
G}{\bigg[}\frac{(2-J^2)H^2}{2{H_0}^2}-\frac{1}{2}{\Omega_{0m}}(1+z)^3-\frac{(2-J^2)(1+z)H}{6{H_0}^2}\frac{dH}{dz}{\bigg]},
\eqno(27)
$$
$$
(\frac{d\phi}{dz})^2=\frac{1}{\lambda}\frac{3{H_0}^2}{8\pi
G}{\bigg[}\frac{2}{3{H_0}^2(1+z)H}\frac{dH}{dz}-\frac{{\Omega_{0m}}(1+z)}{H^2}{\bigg]},\eqno(28)
$$
where $\frac{dz}{dt}=-(1+z)H(z)$. And the equation of state for
dark energy is
$$
\omega^{'}(z)=\frac{p^{'}}{\rho^{'}}=\frac{\frac{(4-J^2)(1+z)}{3(2-J^2)H}\frac{dH}{dz}-1}{1-\frac{2H_0^2}{(2-J^2)H^2}{\Omega_{0m}}(1+z)^3+\frac{J^2(1+z)}{3(2-J^2)H}\frac{dH}{dz}}.\eqno(29)
$$
Eqs.(27)-(29) show that potential $V(z)$ and equation of state
$\omega^{'}(z)$ are only dependent of double imaginary unit $J$,
but field function $\phi$ is dependent of $\lambda$, i.e. $V(z)$
and $\omega^{'}(z)$ are model-independent. Therefore, the
following quantities are shown for $\lambda=1$ (ordinary scalar
field). When $J=i$ (corresponding to the ordinary complex 4D
spacetime), Eqs.(27)-(29) are potential $V(z)$, field $\phi$ and
equation of state $\omega^{'}(z)$ in the ordinary complex
symmetric gravitational theory (OCSGT). When $J=\varepsilon$
(corresponding to the hyperbolic complex 4D spacetime), $V(z)$,
$\phi$ and $\omega^{'}(z)$ are obtained in the
hyperbolic complex symmetric gravitational theory (HCSGT)$^{[14]}$.\\
\mbox{}\hspace{15pt}In the following we will concretely study the
state parameters $\omega^{'}(z)$ corresponding to
$J=i,\varepsilon$ from the observational data. Hence we will deny
the equation of state $\omega^{'}(z)$ in the HCSGT
($J=\varepsilon$) for the constraint condition $p_J=0$. The
observational data we will adopt are the Full Gold dataset (FG)
(157 data points $0<z<1.7$) compiled by Riess et al$^{[17]}$,
which is one of the most
reliable and robust SnIa datasets existing. And all reconstructed quantities are shown for $\Omega_{m}=0.3$.\\
\mbox{}\hspace{15pt}In spatially flat cosmology, the luminosity
distance $D_L(z)$ and Hubble parameter $H(z)$ are simply related
as $(c=1)$
$$
H(z)\equiv\frac{\dot{a}}{a}={\bigg[}\frac{d}{dz}{\bigg(}\frac{D_L(z)}{1+z}{\bigg)}{\bigg]}^{-1}.\eqno(30)
$$
\mbox{}\hspace{15pt}We use a rational ansatz for the luminosity
distance $D_L(z)$$^{[11]}$
$$
\frac{D_L}{1+z}\equiv\frac{2}{H_0}{\bigg[}\frac{z-\alpha\sqrt{1+z}+\alpha}{\beta
z+\gamma\sqrt{1+z}+2-\alpha-\gamma}{\bigg]},\eqno(31)
$$
where $\alpha,\beta$ and $\gamma$ are fitting parameters, which
are determined by minimizing the function
$$
\chi^2(a_1...a_n)=\sum_{i=1}^{N}\frac{(\mu_{obs}(z_i)-\mu_{th}(z_i))^2}{\sigma_i^2},\eqno(32)
$$
where the total error published for the FG dataset
$\sigma_i^2=\sigma_{\mu i}^2+\sigma_{int}^2+\sigma_{\nu i}^2$. The
observational and theoretical distance modulus are defined as
$\mu_{obs}(z_i)=m_{obs}(z_i)-M$ and
$\mu_{th}(z_i)=5\log_{10}(D_L(z))+\mu_0$, respectively. The
luminosity distance $D_L$ is related to the measured quantity, the
corrected apparent peak $B$ magnitude $m_B$ as
$m_B=M+25+5\log_{10}D_L$, where $M$ is the absolute peak
luminosity. The minimization of Eq.(32) is made using the
FindMinimum command of Mathematica. Moreover, the following
constraints$^{[11]}$ are applied to minimize $\chi^2$,
$$
\Omega_m=(\frac{\beta^2}{\alpha\beta+\gamma})^2,\eqno(33)$$
and
$$ \frac{4\beta+2\gamma-\alpha}{2-\alpha}\geq
3\Omega_m.\eqno(34)$$
\mbox{}\hspace{15pt}Our reconstructions for
$\omega^{'}(z)$ are shown in fig1.(The curves plotted are for the
best-fit values of the parameters
$\alpha=1.492,\beta=0.583,\gamma=-0.190$) Notice that the
divergence of equation of state $\omega^{'}(z) (J=\varepsilon)$ in
the hyperbolic complex spacetime does not accord with observation
facts. So we deny this situation. According to the evolution of
$\omega^{'}(z) (J=i)$ in the OCSGT, we find that the
$\omega^{'}(z)$ remains close to -1 at present ($z=0$) and the
fate of the universe would be Big Rip in the future. It is
interesting that $\omega^{'}(z)$ changes from below -1 to above
-1, which agrees with the Quintom model$^{[9,10]}$. In fig 2 and
3, the potential $V(z)$ reconstructed and the age of the universe
obtained using $ t(z)=H_0^{-1}\int^\infty_z\frac{H_0dz}{(1+z)H}$,
are respectively shown. And we see that the age of universe is
about 14 Gyr, which is consistent with the observations.
\begin{figure}
\includegraphics{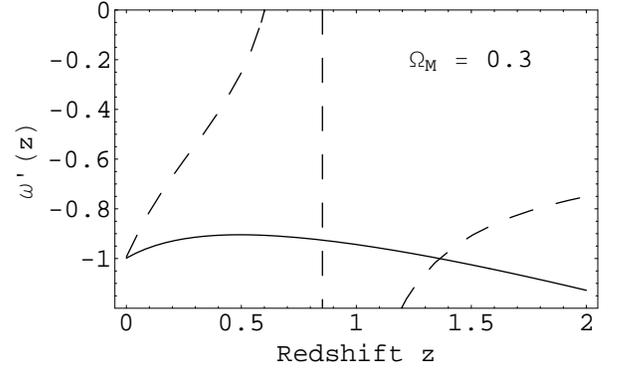}
\caption{The plot shows the evolution of $\omega^{'}(z)$ for the
$\Omega_m=0.3$ in the DCSGT. The solid line  and the dashed line
respectively denote the $\omega^{'}(z)$ in the OCSGT $(J=i)$ and
HCSGT $(J=\varepsilon)$. Notice that the divergence of
$\omega^{'}(z)$ in the HCSGT is not consistent with the
observations while the $\omega^{'}(z)$ accords with the
observations in the OCSGT and changes from below $-1$ to above
$-1$.} \label{fig1}
\end{figure}
\begin{figure}
\includegraphics{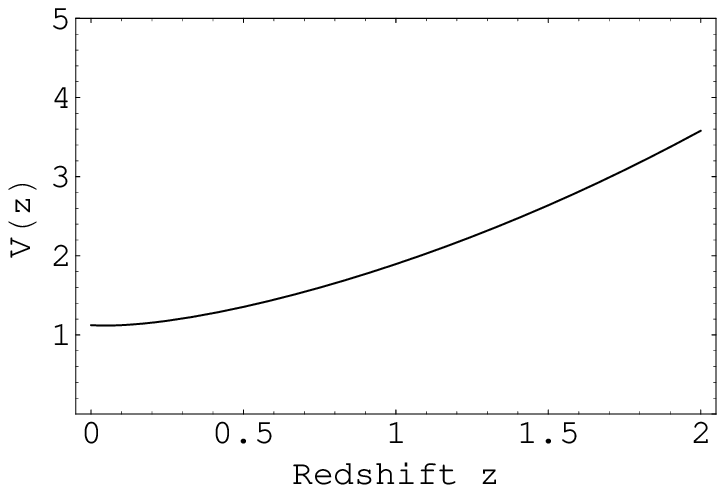}
\caption{The effective potential $V(z)$ is shown in units of
$3H_0^2/8\pi G$.} \label{fig2}
\end{figure}
\begin{figure}
\includegraphics{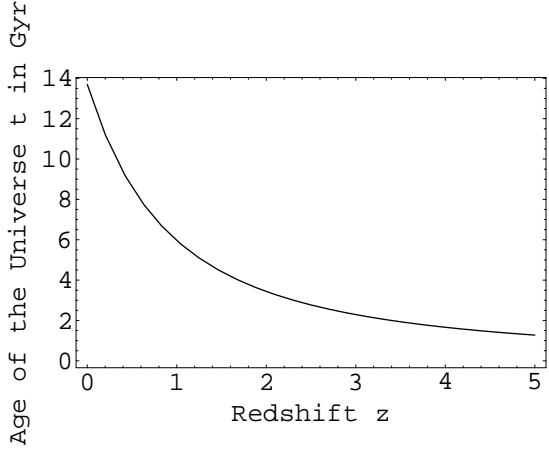}
\caption{The age of the Universe at a redshift $z$, given in Gyr,
for the value of $H_0=70km s^{-1} Mpc^{-1}$.} \label{fig3}
\end{figure}

\subsection{the case of $J^2=1$}
We have discussed the case of $p_J=0, J=\varepsilon$ above. Below
we will study the case of $J^2=1 (J=\varepsilon)$, but $p_J\neq
0$. Eq.(24) turns into
$$
H^2=\frac{8\pi
G}{3}{\bigg[}\frac{5}{6}{\rho_m}+\frac{\lambda}{2}\dot{\phi}^2+\frac{2}{3}V(\phi)+\frac{2}{3}{\rho_J}{\bigg]}.\eqno(35)
$$
\mbox{}\hspace{15pt}If the density parameter $\Omega_J$ in the
imaginary part of the whole spacetime ${\cal M}^4_C(J)$ is $
\Omega_J\equiv\frac{\rho_J}{\rho_0}=\rho_J/(\frac{3H_0^2}{8\pi
G})=1, $ then we can obtain
$$
V(z)=\frac{3H_0^2}{8\pi
G}{\bigg[}\frac{3H^2}{2H_0^2}-\frac{1}{2}\Omega_{0m}(1+z)^3-\frac{(1+z)H}{2H_0^2}\frac{dH}{dz}-1{\bigg]},\eqno(36)
$$
$$
(\frac{d\phi}{dz})^2=\frac{1}{\lambda}\frac{3H_0^2}{8\pi
G}{\bigg[}\frac{2}{3H_0^2(1+z)H}\frac{dH}{dz}-\frac{1}{H^2}\Omega_{0m}(1+z){\bigg]}.\eqno(37)
$$
Furthermore, the equation of state for dark energy
$$
\omega^{'}(z)=\frac{p^{'}}{\rho^{'}}=\frac{-1+\frac{5(1+z)}{9H}\frac{dH}{dz}+\frac{2H_0^2}{3H^2}}{1-\frac{2H_0^2}{3H^2}{\Omega_{0m}}(1+z)^3-\frac{1+z}{9H}\frac{dH}{dz}-\frac{2H_0^2
}{3H^2}}.\eqno(38)
$$
Eqs. (36)-(38) are potential $V(\phi)$, field $\phi$ and equation
of state $\omega^{'}(z)$ in the hyperbolic complex symmetric
gravitational theory (HCSGT). We still apply the $D_L(z)$ ansatz
(31) and the two constraints (33) and (34) to minimize $\chi^2$.
Hence the equations of state $\omega^{'}(z)$ and potential $V(z)$
for dark energy are shown in fig4 and 5, respectively. Note that
the $\omega^{'}(z)$ remains close to -1 at present and changes
from below -1 to above -1 for the whole redshift range $0<z<1.7$.
The fate of the universe would be Big Rip in the future.
\begin{figure}
\centering
\includegraphics{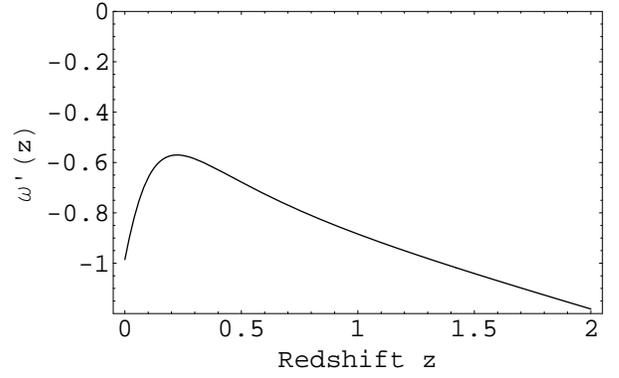}
\caption{The equation of state $\omega^{'}(z)=p^{'}/\rho^{'}$ as a
function of redshift $z$ for $p_J=0$ in the HCSGT
($J=\varepsilon$). The curve plotted is for $\Omega_m=0.3$. The
$\omega^{'}(z)$ changes from below -1 to above -1 and remains
close to -1 at present.}\label{fig4}
\end{figure}
\begin{figure}
\includegraphics{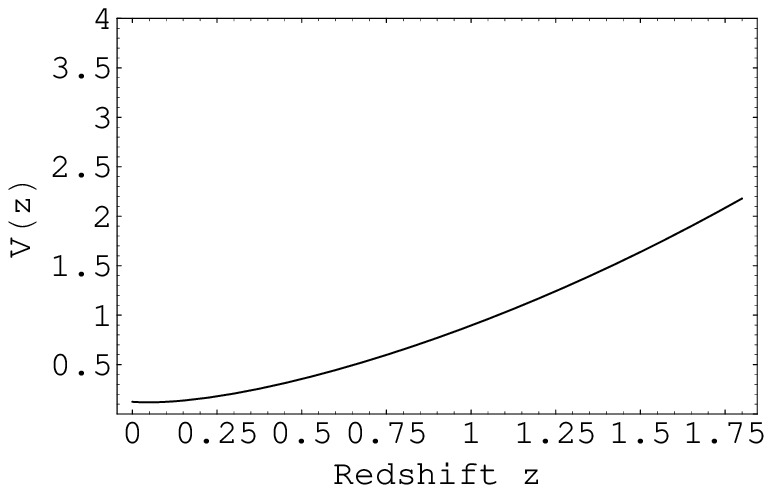}
\caption{The effective potential $V(z)$ is shown in units of
$3H_0^2/8\pi G$.} \label{fig5}
\end{figure}
\section{CONCLUSIONS}
In this paper, we have proposed to study the accelerating
expansion of the universe in the double complex symmetric
gravitational theory (DCSGT). The universe we live in is taken as
the real part of the whole spacetime ${\cal M}^4_C(J)$ which is
double complex. By introducing the spatially flat FRW metric, the
double Friedmann Equations and two constraint conditions $p_J=0$
and $J^2=1$ have been obtained. Furthermore, using parametric
$D_L(z)$ ansatz and real observational data, we have reconstructed
the equation of state $\omega^{'}(z)$ and potential $V(\phi)$ for
the two constraints, respectively. The results have indicated that
the corresponding state parameters $\omega^{'}(z)$ are consistent
with the observations for the two cases of $J=i,p_J=0$ and $J^2=1
(J=\varepsilon),p_J\neq 0$. So we have concluded that the whole
spacetime ${\cal M}^4_C(J)$, may be either ordinary complex
($J=i$) for $p_J=0$ or hyperbolic complex ($J=\varepsilon
(J^2=1)$) for $ p_J\neq 0$. Moreover, we find (see fig1 and 4)
that the $\omega^{'}(z)$ remains close to the $-1$ at present
($z=0$) and changes from below -1 to above -1, which is consistent
with the Quintom model$^{[9,10]}$. And fig1 and 4 show that when
$z<0, \omega^{'}(z)<-1$ which tell us that the fate of the
universe would be Big Rip in the future. Since we have only
studied the corresponding properties of the dark energy in the
DCSGT, some questions are not deeply discussed (eg. the relation
(16) of matter between real and imaginary space and the properties
of the whole spacetime). Hence, we will investigate these issues
in detail in the forthcoming work.
\section{ACKNOWLEDGEMENTS}
This work was supported by National Science Foundation of China
under Grant NO.10573004.

\end{document}